\documentclass[prd,superscriptaddress,nofootinbib,showpacs,preprint]{revtex4}
\usepackage{graphicx}
\usepackage[usenames,dvipsnames]{color}
\usepackage{amsmath,amssymb}
\usepackage[colorlinks]{hyperref}
\newcommand{\vev}[1]{\langle #1 \rangle}
\begin{document}
\vspace*{2cm}
\title{Testable Leptogenesis in extended Standard Model}
\author{\textcolor{blue}{Sudhanwa Patra}}
\email{sudhakar@prl.res.in}
\affiliation{\textcolor{blue}{Physical Research Laboratory, Ahmedabad-380009, India}}
\begin{abstract}

We have proposed a new minimal extension of the Standard Model with a heavy Majorana 
fermion triplet($\Sigma$) and an extra scalar doublet($\eta$) so that the seesaw mechanism is radiative and can 
be accessible at upcoming accelerators. The origin of neutrino mass via the famous seesaw mechanism through 
the heavy Majorana fermion triplet has been discussed. We have proposed a mechanism of leptogenesis by the 
decay of the lightest neutral component of the fermion triplet into a Standard Model lepton doublet and an extra Higgs 
doublet. The important thing is that the leptogenesis scenario discussed in this letter can be of TeV scale 
and hence can be testable at Large Hadron Collider(LHC). We have also discussed a possible dark matter 
scenario in our model and the possible phenomenology of the fermion field $\Sigma$.
\end{abstract}

\maketitle
\section{Introduction}
\label{sec:intro}
Neutrino mass is by far the most important subject of study in neutrino physics
and the mass of neutrinos has been the topic of intense experimental and theoretical 
investigation. There is convincing evidence of neutrino mass from several experiments 
and the landscape of particle physics has been fundamentally altered forever by this 
discovery. The theoretical question is not only how one can extend the Standard Model(SM) to find 
models with massive neutrinos, but also how one can understand the smallness of the neutrino 
mass compared to that of the charged fermions. The origin of the small neutrino mass is still a mystery. 
It is commonly believed that neutrino masses are a low-energy manifestation of physics beyond 
the Standard Model and that their smallness is due to a suppression generated by a new high-energy scale, 
perhaps related to the unification of forces. This is achieved, for example, with the 
celebrated seesaw mechanism \cite{minkowski1977,yana1979,grs1979,glashow1980,ms1980,sv1980} involving heavy particles.There are 
three different kinds of heavy particles that can induce an effective operator $L=y_{eff} 
\dfrac{\ell \ell \phi \phi}{M}$ (where $\ell$ is the SM lepton doublet,
$\phi$ is the SM Higgs doublet) 
which can provide small neutrino mass for$M \gg M_{W}$. There are three kinds of seesaw mechanism occurs in nature 
via the following three kinds of heavy particle: 
\begin{enumerate}
\item Standard Model fermion singlet, coupled to the leptons through Dirac Yukawa couplings and 
usually called right handed neutrinos (type I seesaw \cite{yana1979,grs1979,sv1980,ms1980}).
\item Triplet scalar under $SU(2)$ with Y=2 coupled to leptons through Dirac Yukawa couplings (type II seesaw 
\cite{grimus,lidner}).
\item SU(2) fermion triplet (Y=0) coupled to leptons through Dirac Yukawas (type III seesaw \cite{joshi, barr1}).
\end{enumerate}
One assumes that there were equal amounts of matter and antimatter in
the Universe at the time of the 
Big Bang or, at the end of reheat after inflation.  But we observe today
an enormous preponderance of 
matter compared with antimatter \cite{kt1990}. Leptogenesis has been extensively studied in first and 
2nd kind of seesaw(for right handed singlet fermion and triplet scalar case respectively)\cite{sv1980,sv2006,bdp2003,mmm2009,ms1999,fy1986,ry1986,luty1992,
mz1992,fps1995,fpsw1996, reslepto} which relates neutrino properties and hence the requirement of successful baryogenesis 
yields stringent constraint on the neutrino mass(both light and heavy neutrino case). 

A major part of matter in the Universe is not visible and much of this is Dark Matter. 
Therefore we can define two categories: Baryonic Dark Matter, composed of baryons which are not 
seen, and Non-Baryonic Dark Matter, composed of massive neutrinos, or elementary particles which 
are as yet undiscovered. The particles which comprise non-baryonic dark matter must have survived
from the Big Bang or, at the end reheat after inflation and therefore must be stable or have lifetimes in excess of the current age
of the Universe. Recent cosmological observations \cite{wmap2003} not only tell us how much dark matter exists 
but also that it must be nonbaryonic– it is not one of the known elementary particles contained 
within the Standard Model of particle physics. Dark matter is a known unknown. We do not know what 
the underlying theory of dark matter is, what are the detailed particle properties of it, nor the 
particle spectrum of the dark sector.

In this paper, we have discussed the leptogenesis scenario by the decay of the lightest neutral component 
of fermion triplet and it's phenomenological implications which may be testable in near future. The paper is 
organized like this: Section (\ref{sec:ourmodel}) give the particle content of an extra fermion triplet per 
each generation and an extra Higgs doublet which are different from Standard Model particles by a discrete 
symmetry($Z_{2}$). In section (\ref{sec:neumass}) we discuss the neutrino mass generation by radiative mechanism. 
In section (\ref{sec:lepto}), the kinematical effect of the leptogenesis have been discussed and the scenario 
can alter the out-of equilibrium condition substantially. Here we gave the first full calculation of thermal 
leptogenesis by the decay of $SU(2)$ triplet fermion $\Sigma$. Section (\ref{sec:darkmatter}) discusses possible 
dark matter candidate in our model and discussed the possible phenomenology of fermion triplet $\Sigma$.
Finally, section (\ref{sec:conclusion}) gives the conclusion.
\section{Our Model}
\label{sec:ourmodel}
In our model we extend the Standard Model by introducing a three generation of fermionic triplet neutrinos 
$\Sigma_{i}$(i=1,2,3) and one more Higgs doublet $\eta$. We impose a $Z_{2}$ discrete symmetry on this model. 
Under this symmetry transformation $\Sigma_{i}$ and $\eta$ change sign, while other fields remain 
the same. As a result of this symmetry, $\Sigma_{i}$ does not couple to the standard Higgs $\phi$ 
through Yukawa coupling. Only the new Higgs doublet $\eta$ couples to $\Sigma$ . 
\begin{center}
\begin{tabular}{|c|cccccccc|}
\hline 
 & $Q_{\alpha}$ & $u^{c}_{\alpha}$ & $d^{c}_{\alpha}$ & $\ell_{\alpha}$ & $e^{c}_{\alpha}$  
& $\phi$ & $\Sigma$ & $\eta$  \\
\hline 
$Z_{2}$ & +1 & +1 & +1 & +1 & +1 & +1 & -1  & -1  \\
\hline 
\end{tabular}
\end{center}
Now we can write down all the possible interaction terms for this model. It includes the
gauge interactions, Yukawa couplings and the Higgs potential. However in this work only the 
Yukawa interaction for the lepton and part of the Higgs potential are relevant. The Yukawa 
coupling and complete Higgs potential are written as
\begin{eqnarray}
{\cal L}_Y&=&\sum_{\alpha,i=e,\mu,\tau} \left(f_{\alpha i} \bar{\ell}_{\alpha} \phi \,e_{Ri} +
h_{\alpha i}\bar{\ell}_{\alpha} \,i \tau_{2} \Sigma_{i} \eta \right)+
\frac{1}{2} M_{\Sigma} \bar{\Sigma}^{c} \Sigma+ {\rm h.c.}. 
\label{lag}
\end{eqnarray}
and
\begin{eqnarray}
V&=&\frac{1}{2}\lambda_1(\phi^\dagger \phi)^2 +\frac{1}{2} \lambda_2(\eta^\dagger\eta)^2
+\lambda_3(\phi^\dagger \phi)(\eta^\dagger\eta)   \nonumber \\
&+&\lambda_4(\phi^\dagger\eta)(\eta^\dagger \phi)+m_\phi^2 \phi^\dagger \phi+m_\eta^2 \eta^\dagger\eta 
\nonumber \\
&+&\frac{1}{2} \lambda \left[ (\phi^{\dagger} \eta)^{2} +(\eta^{\dagger} \phi)^{2}\right]
\label{pot}
\end{eqnarray}
Here $\ell_{\alpha}$ and $e_{Ri}$ are the lepton doublet and the right-handed charged lepton respectively. 
Since $Z_{2}$ symmetry is exact and will not be broken, $\eta(\eta^{0},~\eta^{-})$  will not develop a non-zero vacuum 
expectation value. Therefore the neutrino does not obtain mass at tree level as lepton number is conserved. 
However, given a term like $\frac{1}{2} \lambda \left[ (\phi^{\dagger} \eta)^{2} +(\eta^{\dagger} \phi)^{2}\right]$ 
in the potential, lepton number is not automatically conserved, but the lepton number violation should be small. 
Hence, the value of the coupling $\lambda$ must be small. In other words, neutrinos gets a non-zero mass, 
but obviously this mass is generated only at the loop level. For convenience and without loss of generality, 
we will chose the basis where the fermion triplet mass matrix is real and diagonal.
\section{Neutrino mass by Fermion triplet}
\label{sec:neumass}
In this section, the generic type III seesaw mechanism by fermion triplet has been discussed. However, we 
don't interest in generic seesaw. Then we have shown that how samllness mass came from radiatively via 
triplet fermions at one loop diagram.

 For generic type III seesaw mechanism, one must replace three right handed singlet neutrinos by three fermion 
triplet neutrino under SU(2). The fermionic triplet is given as. 
\begin{equation}\label{eq1:triplet_sigma_2rep}
 \Sigma =\left( \begin{tabular}{cc}
 $\Sigma^{-}$        & $\Sigma^0/\sqrt{2}$   \\
 $\Sigma^0/\sqrt{2}$ & $\Sigma^+$      
          \end{tabular}\right).
\end{equation} 
\begin{figure}[ht]
 \centering
 \includegraphics[bb=162 620 433 721]{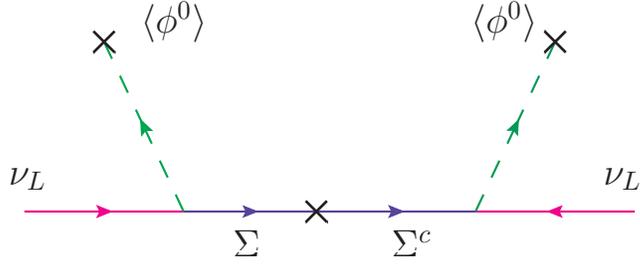}
\caption{Type III seesaw with heavy fermion triplet}
\label{neumass:triplet}
\end{figure}
The corresponding Lagrangian for this model is given 
\begin{equation}\label{eq1:L_mass_type3}
 -\mathcal{L}_\text{type-III} =
  h_{\alpha i} \, \overline{\ell}_L\, i\tau_2\, \Sigma \,\phi
  + \frac{1}{2} M_\Sigma\, \text{Tr}\left(\overline{\Sigma^c}\, \Sigma\right)
  +\text{h.c.}\;
\end{equation} 

This gives rise to the diagram shown in Fig.(\ref{neumass:triplet}), and after integrating out 
the heavy $\Sigma$ field, one obtains the desired form for the seesaw neutrino mass
\begin{equation}
 m_\text{eff}^{\rm III} \simeq \,
  h_{\alpha i}\,\frac{\vev{\phi^0}^2}{M_\Sigma}\,h_{\alpha i}^T
  \;.
  \label{eq1:type3_mass_eff}
\end{equation}
Hence by setting $M_\Sigma \gg \vev{\phi^0}$, one can explain the smallness of neutrino masses. 
This is often referred to as the type III seesaw mechanism .But this is given for 
general feelings of type III seesaw to generate small neutrino mass. But we have different approach 
to get neutrino mass radiatively.Then the above Yukawa term is not allowed in our model. Neutrino mass 
generation radiatively will be discussed in next subsection.
\subsection{Neutrino mass by loop effect}
There are no Dirac masses between the left and right-handed
neutrinos since the exact $Z_{2}^{}$ symmetry protects $\eta$ from obtaining 
vacuum expectation values. So, the neutrinos will remain
massless at tree level unless the quartic scalar interaction
\begin{eqnarray}
\label{quartic} \mathcal{L}\supset
-\frac{1}{2}\lambda\left[\left(\phi^{\dagger}_{}\eta\right)^{2}_{}+\textrm{h.c.}\right]\,,
\end{eqnarray}
is included which gives mass at the one loop level as shown in Fig:2. 
Because of the assumed $Z_2$ symmetry, $\nu_\alpha$ does not couple to $\Sigma_i$ 
through $\phi^0$.  Hence $\Sigma_i$ is not the Dirac mass partner of $\nu_\alpha$ 
as in the canonical seesaw model \cite{mmm2009,m2005,mm2009,ma1998}.  Instead, $\nu_\alpha$ couples to $\Sigma_i$ 
through $\eta^0$ (which has no vacuum expectation value) and obtains a 
radiative Majorana mass in one loop \cite{raidal}, i.e. 

\begin{center}
\begin{figure}[t]
 \includegraphics[width=8cm, height=4cm]{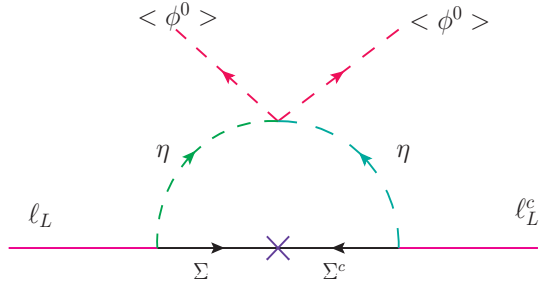}
\caption{The one loop diagram for light neutrino mass generation 
via radiative mechanism through \hspace{1.5cm}fermion triplet}
\end{figure}
\end{center}
\begin{eqnarray}
&&{\cal M}_{\alpha\beta}=\sum_i\Lambda_i h_{\alpha i}h_{\beta i},
 \nonumber \\
&&\Lambda_i=\frac{\lambda \langle\phi\rangle^2}{8\pi^2 M_{\Sigma_i}}~
I\left(\frac{M_{\Sigma_i}^2}{M_\eta^2}\right), \qquad
I(x)=\frac{x}{1-x}
\left(1+\frac{x~\ln~x}{1-x}\right),
\label{nmass}
\end{eqnarray}
where $M_\eta^2=m_\eta^2+(\lambda_3+\lambda_4)\langle\phi\rangle^2$.
This neutrino mass matrix can explain the neutrino oscillation data well
as long as we set appropriate values for the parameters 
$\lambda$, $h_{\alpha i}$, $M_{\Sigma_i}$ and $M_\eta$.
We note that $\lambda$ should be very small to generate desired neutrino
masses. However, this tuning is not so bad nature since it can be
controlled by a global symmetry which appears if we make $\lambda$
zero.

For $x_i >> 1$, i.e. $\Sigma_i$ very heavy,
\begin{equation}
({\cal M}_\nu)_{\alpha \beta} = \frac{\lambda v^2 }{ 8 \pi^2} \sum_i 
\frac{h_{\alpha i} h_{\beta i} }{ M_i} [\ln x_i - 1]
\end{equation}
instead of the canonical seesaw expression of $v^2 \sum_i h_{\alpha i} h_{\beta i} 
/M_i$. Therefore, the observed neutrino mass can be naturally explained. For example, 
a value of $m_{\nu}$ can be obtained for a choice of parameters like 
$\lambda \sim {\cal O}(10^{-4})$, $h\sim {\cal O}(10^{-3})$ and $M_{\Sigma} 
\sim {\cal O}(\text{1000 GeV to 10 TeV})$. 

To check the consistency of this model with observed results on neutrino mass, 
we present a particular flavor structure of the Yukawa sector which fits the 
neutrino data appropriately. Consider such flavor structure for neutrino Yukawa 
couplings as 
\begin{equation}
h_{ei}=0, \quad h_{\mu i}=h_{\tau i}~ (i=1,2); \quad 
h_{e3}=-h_{\tau 3}, \quad \quad h_{\mu 3}=-h_{\tau 3}.
\label{Yukawa}
\end{equation}
In this case the neutrino mass matrix can be written as
\begin{equation}
{\cal M}=\left(\begin{array}{ccc}
0 & 0 & 0\\ 0 & 1 & 1 \\ 0 & 1 & 1 \\
\end{array}\right)(h_{\tau 1}^2\Lambda_1+h_{\tau 2}^2\Lambda_2)+
\left(\begin{array}{ccc}
1 & 1 & -1\\ 1 & 1 & -1 \\ -1 & -1 & 1 \\
\end{array}\right)h_{\tau 3}^2\Lambda_3,
\end{equation}
and the tri-bimaximal neutrino mixing is automatically
realized for the neutrino mass matrix (\ref{nmass}) \cite{sty,tex}. 
However, only two mass eigenvalues take nonzero values.
Thus, the neutrino oscillation data can be consistently explained 
as long as the following conditions are satisfied: 
\begin{equation}
h_{\tau 1}^2\Lambda_1+h_{\tau 2}^2\Lambda_2
\simeq 2.5\times 10^{-2}~{\rm eV}, \qquad
h_{\tau 3}^2\Lambda_3\simeq 2.9\times 10^{-3}~{\rm eV}. 
\label{coscil}
\end{equation}
These come from the required values for $\Delta m_{\rm atm}^2$ and
$\Delta m_{\rm solar}^2$, respectively.
We need to consider the constraints from both the lepton flavor violating 
processes. In leptogenesis, the lightest $M_{\Sigma_i}$ may then be much below the 
Davidson-Ibarra bound \cite{bdp2003,di2002} of about $10^9$ GeV, thus avoiding a 
potential conflict of gravitino overproduction and thermal leptogenesis. 
In this scenario, $\eta^0$ is dark matter.
\section{Leptogenesis}
\label{sec:lepto}
Leptogenesis provides a simple and elegant explanation of the cosmological matter-antimatter 
asymmetry. A beautiful aspect of this mechanism is the connection between the baryon asymmetry 
and neutrino properties. Just as there are three seesaw mechanisms, the decays of the 
corresponding heavy particles N \cite{fy1986}, ($\Delta^{++}$, $\Delta^{+}$, $\Delta^{0}$) 
\cite{ms1999}, and ($\Sigma^{+}$, $\Sigma^{0}$, $\Sigma^{-}$ ) \cite{hambye2006,f2008} are natural for 
generating a lepton asymmetry of the Universe, which gets converted \cite{k1985} into the 
present observed baryon asymmetry through anomalous baryon number violating interactions. Just as N may decay into leptons and anti leptons
because it is a Majorana fermion, the same is true for $\Sigma$. The heavy Majorana neutrinos 
$\Sigma=\Sigma+\Sigma^{c}$ decays to SM left-handed doublet leptons and doublet scalars. 
The decay channel $\Sigma^{0}_{1} \rightarrow \ell_{L} \eta$ and its CP conjugate decay channel 
$\Sigma^{0}_{1} \rightarrow \ell^{c}_{L} \eta$ contribute to the lepton asymmetry. The Lepton asymmetry parameter 
can be calculated by the interference of the decay of $\Sigma^{0}_{1}$ at tree and one loop diagram.

Leptogenesis is expected to occur through the decay of $\Sigma^{0}_{1}$ into a light lepton and an 
extra Higgs doublet. If departure from equilibrium in $\Sigma^{0}_{1}$ decay is large, the lepton 
asymmetry is
\begin{equation}
\label{nl}
n_L/s \simeq \frac{\varepsilon^{\Sigma}}{s} \frac{g_{\Sigma} T^{3}}{\pi^{2}}
\end{equation}
where $g_{\Sigma}=2$ is the spin degrees of freedom of fermion triplet component., so that their 
number density before they decay is $\sim g_{\Sigma} T^{3}/\pi^{2}$ if we assume they go out of 
equilibrium before they become non-relativistic. Using $s=(2/45) g_{*} \pi^{2} T^{3}$, where 
$g_{*}=g_{\text{boson}}+(7/8)g_{\text{fermions}}$ is the effective relativistic degree of freedom 
contributing to the entropy($g_{*}=106.75$), we get
$$n_L/s \simeq 4 \times 10^{-3} \varepsilon^{\Sigma}.$$
This lepton asymmetry will then be reprocessed by anomalous electroweak sphaleron process, leading 
to a baryon asymmetry 
\begin{equation}
 n_B/s=\left( \frac{\text{24+4} N_{\phi}}{\text{66+13} N_{\phi}} \right) n_{B-L}/s \simeq 
-\frac{28}{79} n_L/ s. 
\label{nb}
\end{equation}
Combining equations (\ref{nl}) and (\ref{nb}), we get the present baryon asymmetry as
\begin{equation}
n_B/ s \simeq -4 \times \frac{28}{79} 10^{-3}\varepsilon^{\Sigma}
\end{equation}
Values of $\varepsilon^{\Sigma} \sim 10^{-7}$ is required to account for the value 
$\eta_{B}=n_B/n_{\gamma}  \simeq (2-6) \times 10^{-10}$ obtained from CMBR result 
by WMAP\cite{wmap2003} and from the successful theory of primordial nucleosynthesis.

Leptogenesis of first kind(by a right handed singlet fermion) was first proposed by Fukugita and Yanagida 
\cite{fy1986}. But leptogenesis of 2nd and 3rd kind(by a scalar triplet and a 
fermion triplet) is not so simple. The main concern with the other possibilities is that gauge
scatterings can keep $SU(2)_L$ triplets close to thermal equilibrium, conflicting with the 
third Sakharov condition. In this Letter, we carried out the calculation of the lepton asymmetry for the fermion 
triplet case and have discussed possible issues related to it.
In our model, the decay of the heavy fermionic triplet neutrino(but lightest among the 
three generation and neutral i.e, $\Sigma^{0}_{1}$) can create a lepton asymmetry, if the Yukawa couplings $h$
provide the source of the CP violation. The lowest order non-trivial
asymmetry comes from the interference of the tree-level diagrams
with the one-loop diagrams. We calculate the CP asymmetry and find
it to be the same as that in the canonical seesaw model. If the fermionic triplet is heavier than the 
extra Higgs field $\eta$, it can decay to a light lepton($\ell_{L}$) and an extra Higgs ($\eta$). 
Using the Yukawa term $h_{\alpha i} \bar{\ell}_{\alpha} (i \tau_{2}) \Sigma_{i} \eta$, 
at the tree level, the decay widths are given by
\begin{eqnarray}
\label{decayneu} &&\Gamma\left(\Sigma_i^{}\rightarrow
\ell_L^{}+\eta^\ast_{}\right)=\Gamma\left(\Sigma_i^{}\rightarrow
\ell_L^{c}+\eta\right)\\
&=&\frac{1}{16\pi}\left(h^{\dagger}_{}h\right)_{ii}^{}\,M_{\Sigma_{i}^{}}^{}r^{2}_{\Sigma_{i}^{}}\,,
\end{eqnarray}
where the factor,
\begin{eqnarray}
\label{kinetics}
r_{\Sigma_{i}^{}}^{}=1-\frac{M_{\eta}^{2}}{M_{\Sigma_{i}^{}}^{2}}\,,
\end{eqnarray}
depends on the mass of the decay product in addition to that of the
decaying particle. For the increasing values of $M_{\eta}^{}$, we
get a smooth interpolation from $r_{\Sigma_{i}^{}}^{}\simeq 1~
(M_{\eta}^{2}\ll M_{\Sigma_{k}^{}}^{2})$ to $r_{\Sigma_{i}^{}}^{}\rightarrow 0
~ (M_{\eta}^{2}\rightarrow M_{\Sigma_{k}^{}}^{2})$. Usually the decay
products are assumed to be very light and hence, this factor is
taken to be $1$. However this factor can also be very small for nearly 
degenerate mass between the $\Sigma$ and $\eta$ field . We
shall consider the case when this factor is very small \cite{gu2008}. In the
present model, the value of parameter $r_{\Sigma_i}$ is taken to be 
very small(of the order of ${\cal O}(10^{-4})$) and has been shown in 
next paragraph how this small value of $r_{\Sigma_i}$ can give required 
CP asymmetry to explain TeV scale leptogenesis. At present we don't have 
any explanation for the smallness of $r_\Sigma$. But we hope that 
when this model will be embedded in any higher theory and some symmetry 
of that theory will provide an explanation.
\begin{figure}[t]
 \centering
 \includegraphics[width=16cm, height=3.5cm]{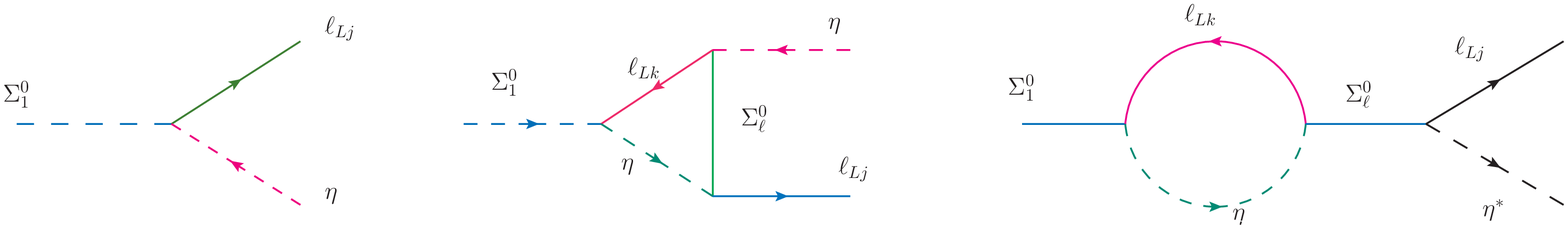}
\caption{decay of Lightest neutral component of the fermion triplet into 
Standard Model lepton doublet and extra higgs doublet (Interference of tree and one loop diagram).}
\end{figure}

The CP asymmetry is given by
\begin{eqnarray}
\label{cpasymmetry} 
\varepsilon_{\Sigma_i}^{}&\equiv & \frac{\Gamma
(\Sigma_i^{}\rightarrow \ell_L^{}+\eta^\ast_{})- \Gamma
(\Sigma_i^{}\rightarrow \ell_L^{c}+\eta)}{\Gamma (\Sigma_i^{}\rightarrow
\ell_L^{}+\eta^\ast_{})+ \Gamma (\Sigma_i^{}\rightarrow
\ell_L^{c}+\eta)}\nonumber\\
&\simeq&\frac{1}{8\pi}\frac{1}{\left(h^{\dagger}_{}h\right)_{ii}^{}}\sum_{j\neq
i}^{}\textrm{Im}\left[\left(h^{\dagger}_{}h\right)_{ij}^{2}\right]\nonumber\\
&&\times
\left[f\left(\frac{M_{\Sigma_{j}^{}}^{2}}{M_{\Sigma_{i}^{}}^{2}}\right)
+g\left(\frac{M_{\Sigma_{j}^{}}^{2}}{M_{\Sigma_{i}^{}}^{2}}\right)\right]\,,
\end{eqnarray}
which is free from the masses of the decay products. Here the
functions $f$ and $g$ are the contributions from the vertex and
self-energy corrections, respectively:
\begin{eqnarray}
\label{vertex}
f(x)&=&\sqrt{x}\left[1-(1+x)\ln\left(\frac{1+x}{x}\right)\right]\,,\\
\label{selfenergy} g(x)&=&\frac{\sqrt{x}}{1-x}\,.
\end{eqnarray}
We consider the case where we have 
$M_{\Sigma_{1}^{}}^{}\ll M_{\Sigma_{2,3}^{}}^{}$ (a factor
$\displaystyle{M_{\Sigma_{2,3}^{}}^{}/M_{\Sigma_{1}^{}}^{}}\displaystyle$ of
$3-10$ is enough because the number density of $\Sigma_{2,3}^{}$ is
rapidly Boltzmann suppressed at the temperature below its mass $M_{\Sigma_{2,3}}$),
where the final lepton asymmetry will mainly come from the
 decay of $\Sigma^{0}_{1}$. We can simplify the lepton
asymmetry (\ref{cpasymmetry}) as
\begin{eqnarray}
\label{cpasymmetry2}
\varepsilon_{\Sigma_1}^{}&\simeq&-\frac{3}{16\pi}\frac{1}{\left(h^{\dagger}_{}h\right)_{11}^{}} 
\sum_{j=2,3}^{}\textrm{Im}\left[\left(h^{\dagger}_{}h\right)_{1j}^{2}\right]
\frac{M_{\Sigma_{1}^{}}^{}}{M_{\Sigma_{j}^{}}^{}}\,.
\end{eqnarray}
We remind that when the lightest particle is a fermion triplet neutrino $\Sigma^{0}_{1}$, 
the lepton-asymmetry from its decay is given by the famous relation \cite{di2002,bdp2003,flantz}
(inserting the neutrino mass in that formula (\ref{cpasymmetry2})),
\begin{eqnarray}
\label{cpasymmetry3} |\varepsilon_{\Sigma_1}^{}|&<&
\frac{3\pi}{\mathcal{O}(\lambda)}\frac{M_{\Sigma_{1}^{}}^{}m_{3}^{}}{v^{2}_{}}|\sin\delta|
\end{eqnarray}
with $m_3^{}$ and $\delta$ are the biggest eigenvalues of the
neutrino mass matrix and the CP phase, respectively. Here we have
assumed the neutrinos to be hierarchical \cite{sv2006}. Clearly, in
the present case, the Davidson Ibarra bound \cite{di2002,bdp2003} is relaxed by a factor of
$16\pi^{2}_{}/\mathcal{O}(\lambda)$.

For effectively creating a lepton asymmetry in the thermal evolution
of the universe, the decays of $\Sigma_{1}^{}$ should satisfy the
condition of departure from equilibrium, which is described by
\begin{eqnarray}
\label{condition} \Gamma_{\Sigma_{1}^{}}^{}\lesssim
H(T)\left|_{T=M_{\Sigma_{1}^{}}^{}}^{}\right.\,.
\end{eqnarray}
where
\begin{eqnarray}
\label{decaywidth2} \Gamma_{\Sigma_{1}^{}}^{}&=&
\Gamma\left(\Sigma_{1}^{}\rightarrow
\ell_L^{}+\eta^\ast_{}\right)+\Gamma\left(\Sigma_{1}^{}\rightarrow
\ell_L^{c}+\eta\right)\nonumber\\
&=&\frac{1}{8\pi}\left(h^{\dagger}_{}h\right)_{11}^{}\,M_{\Sigma_{1}^{}}^{}r^{2}_{\Sigma_{1}^{}}
\end{eqnarray}
is the total decay width of $\Sigma_{1}^{}$ and
\begin{eqnarray}
\label{hubble}
H(T)&=&\left(\frac{8\pi^{3}_{}g_{\ast}^{}}{90}\right)^{\frac{1}{2}}_{}\frac{T^{2}_{}}{M_{\textrm{Pl}}^{}}
\end{eqnarray}
is the Hubble parameter with the Planck mass
$M_{\textrm{Pl}}^{}\simeq 1.2 \times 10^{19}_{}\,\textrm{GeV}$ and the
relativistic degrees of freedom $g_{\ast}^{}\simeq 100$ \cite{kt1990}. 
In order to satisfy the out of equilibrium condition, we should have 
\begin{eqnarray}
 & &\Gamma < H(T=M_{\Sigma_{1}}) \nonumber\\
&\Rightarrow & \frac{1}{8\pi}\left(h^{\dagger}_{}h\right)_{11}^{}\,M_{\Sigma_{1}}^{}r^{2}_{\Sigma_{1}^{}}
< \left(\frac{8\pi^{3}_{}g_{\ast}^{}}{90}\right)^{\frac{1}{2}}_{}\frac{M^{2}_{\Sigma_{1}}}{M_{\textrm{Pl}}^{}}
\nonumber\\
&\Rightarrow & M_{\Sigma_{1}} \geq \frac{M_{Pl}}{10} 
\frac{\left(h^{\dagger}_{}h\right)_{11}^{} \, r^{2}_{\Sigma_{1}^{}}}{8 \pi}
\end{eqnarray}
For $h \sim {\cal O}(10^{-3})$, $r_{\Sigma_{1}^{0}}={\cal O}(10^{-4})$, we found the 
bound on the mass of the fermion triplet as $M_{\Sigma_{1}} \geq 10^{3}$ GeV.

If we take the Yukawa coupling as $h \sim {\cal O}(10^{-3})$, the masses of the neutral 
component of the lightest fermion triplet as $M_{\Sigma^{0}_{1}}={\cal O}(\text{1000 GeV to 
10 TeV})$ and the factor which comes in lepton asymmetry formula as $r_{\Sigma^{0}_{1}}= 
{\cal O}(10^{-4})$, we can get the required value of asymmetry to give consistent 
matter-antimatter asymmetry.
For $\lambda=10^{-4}$, $M_{\Sigma^{0}_{1}}=1$ TeV, and $m_{3}=0.07$ eV and the $\sin \delta=-1$,
we will get
\begin{eqnarray}
\label{baryonasymmetry11}
\frac{n_B^{}}{s}&=&\frac{28}{79}\,\frac{n_{B-L}^{}}{s}=-\frac{28}{79}\,\frac{n_{L}^{}}{s}\nonumber\\
&\simeq&
-\frac{28}{79}\,\varepsilon_{\Sigma_{1}^{}}^{}\frac{n_{\Sigma_{1}^{}}^{eq}}{s}\left|_{T=M_{\Sigma_{1}^{}}^{}}^{}\right.\simeq
-\frac{1}{15}\frac{\varepsilon_{\Sigma_{1}^{}}^{}}{g_\ast^{}}\nonumber\\
&\simeq& 10^{-10}
\end{eqnarray}
as desired to explain the matter-antimatter asymmetry of the
universe.
 
Successful leptogenesis will require that the final result for $\eta_B$ should be order of $10^{-10}$.
These results show that the out-of-equilibrium decay of $\Sigma^{0}_1$ can produce the necessary baryon 
number asymmetry for intermediate values of the mass of the $\Sigma^{0}_1$ as in the usual cases. 
As long as we confine ourselves to the non-supersymmetric framework, the model is free from the 
gravitino problem. If one supersymmetrise the given model, the D-term in 
the Higgs potential($H_u$, $H_d$, $H$ and $H'$ ) will contribute to the neutrino mass via one 
loop diagram. In this scenario, the scale of fermion triplet is around $100$ TeV and hence avoids 
gravitino problem.

The relevant Lagrangian which is important for leptogenesis by fermion triplet is
\begin{eqnarray}
{\cal L}& = &{\cal L}_{\rm SM}+ \bar \Sigma_i i\partial\hspace{-1.3ex}/\, \Sigma_i +
h_{\alpha i}\bar{\ell}_{\alpha} \,i \tau_{2}\,\Sigma_{i} \eta+
\frac{1}{2} M_{\Sigma} \bar{\Sigma}^{c} \Sigma+ {\rm h.c.}) \nonumber \\
&+&\text{gauge interaction part of fermion triplet}
\end{eqnarray}
The gauge interaction term comes from $ D_{\mu} \Sigma D^{\mu} \Sigma$. In this term, 
$ D_{\mu} \Sigma =\partial_{\mu} \Sigma + i g [\tilde{A}_{\mu}, \Sigma]$ and $\tilde{A}_{\mu} =
A^{a}_{\mu} T^{a}$. Here $A^{a}_{\mu}$ and $T^{a}$ are gauge boson and generator of the group.

\begin{figure}[h]
\centering
\includegraphics[width=14cm, height=4cm]{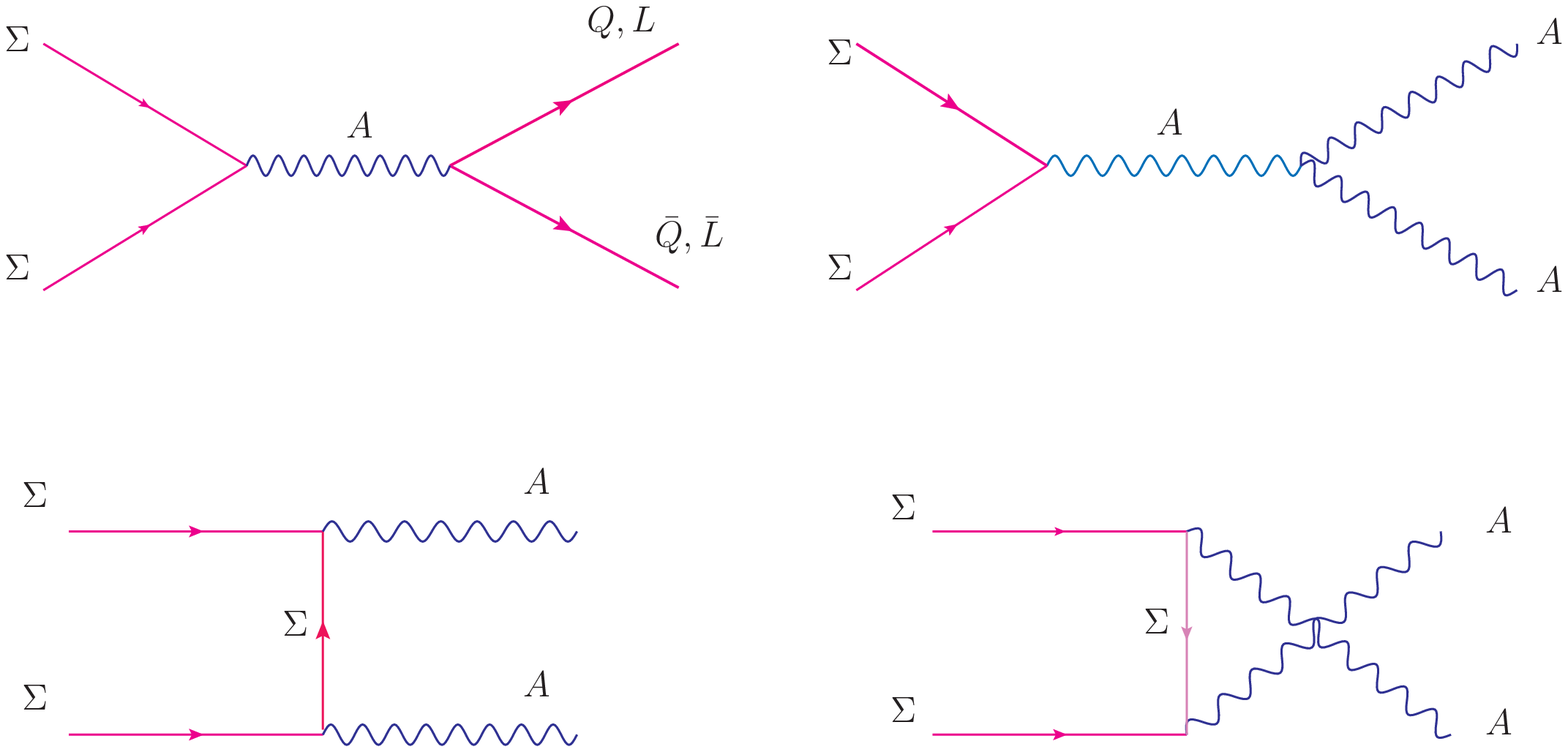}
\caption{Feynman diagrams that contribute to the interaction rate $\gamma_{\Sigma}$}
\end{figure}

Right handed singlet neutrinos trivially allow thermal leptogenesis: as they only have Yukawa interaction, 
but do not have any gauge interaction and hence they easily satisfy the out-of-equilibrium Sakharov 
condition for baryogenesis. Fermion triplets (like scalar triplets) have gauge interactions so that
it is more difficult to have a  non thermal abundance. It's worth mentioning that the gauge interactions 
involve two particles of fermion triplet (see in fig. 4) and are therefore doubly Boltzmann suppressed 
at temperatures below their mass, so that they cannot wash-out the lepton asymmetry in an efficient way. 
On the other hand, gauge interactions are effective at higher temperature and thermalize the initial abundance of $\Sigma^{0}_{1}$, 
but the final baryon asymmetry does not depend on it almost \cite{hambye2004,hambye2006,hambye}.
Presence of gauge interaction try tend to maintain the thermal equilibrium, still can get maximum efficiency 
for any mass even if $M_{\Sigma}=$1 TeV if they satisfy:
(1) If decay rates $\Gamma(\Sigma^{0}_{1} \rightarrow \ell \eta)$ is much faster than the annihilation rate.
(2) Decay rate is slower than the expansion rate of the Universe.
\section{Dark Matter and Phenomenology of $\Sigma$}
\label{sec:darkmatter}
Now we come to discuss the application of our model to the dark matter issue of the Universe and 
other issues in neutrino physics. Here we will restrict ourselves to the dark matter candidate 
that appear from the model of neutrino masses. One natural explanation of neutrino mass is due 
to radiative models. In such models one introduce some additional fermionic and scalar particles 
and impose some extra symmetry. Among these, the lightest neutral component can be candidate 
for dark matter. In this case $\eta$ can be dark matter candidate

If $\Sigma$ mass lies in the range of TeV scale, then we can have clean signature at LHC even if 
only a few events are seen \cite{m2005,hambye2004,ma2006}. The neutral component are seen as missing energy though this is 
difficult to track, but the charged component of $\Sigma$ are enough long-lived so that they 
manifest in the detector an charged tracks. The lifetime of $\Sigma^{\pm}$ particles is 
$\tau \simeq 44/8$ cm and the possible decay channels are 
\begin{equation}
\Gamma\left(\Sigma^{0}\rightarrow \eta +\nu_{k}\right) =
\frac{M_{\Sigma}}{32 \pi } \mid h \mid^{2} \left( 1- \frac{M^{2}_{\eta}}{M^{2}_{\Sigma}}\right)^{2} 
\end{equation}
\begin{figure}[h]
 \centering
 \includegraphics[bb=0 0 240 153]{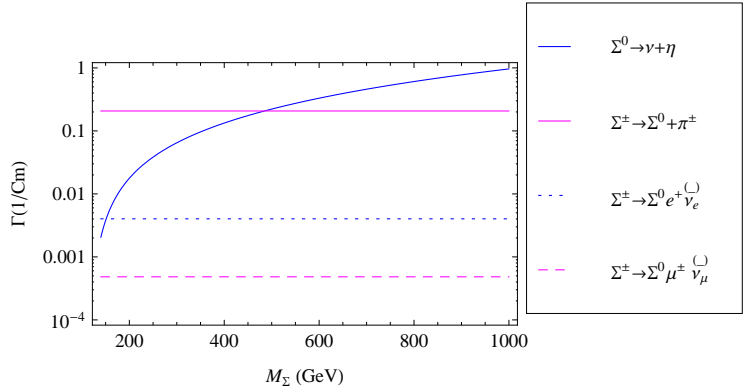}
\caption{Allowed decay rates of $\Sigma$ in our model}
\end{figure}
The mass splitting between charged component and the neutral one is $\sim 167$ MeV. This mass difference is 
bigger than $m_{\pi}$ in all allowed value of $M_{\Sigma}$ and hence, gives extra contribution to the 
following decay channel
\begin{equation}
\Gamma\left(\Sigma^{\pm}\rightarrow \Sigma^{0}+\pi^{\pm}\right) =
\frac{8 G^{2}_{F} V^{2}_{ud} (\Delta M)^{3} f^{2}_{\pi}}{32 \pi } \left( 1- \frac{M^{2}_{\pi}}{(\Delta M)^{2}}\right)
\end{equation}
\begin{equation}
\Gamma\left(\Sigma^{\pm}\rightarrow \Sigma^{0} e^{+} \bar{\nu}_{e},~\Sigma^{0} e^{-} \nu_{e}\right) =
\frac{8 G^{2}_{F} (\Delta M)^{5}}{ 60 \pi^{3} }
\end{equation}
\begin{equation}
\Gamma\left(\Sigma^{\pm}\rightarrow \Sigma^{0} \mu^{+} \bar{\nu}_{e},~\Sigma^{0} \mu^{-} \nu_{e}\right) =
0.12 \times \frac{8 G^{2}_{F} (\Delta M)^{5}}{ 60 \pi^{3} } 
\end{equation}
Where $f_{\pi}=131$ MeV and $\Delta M$=167 MeV. The lifetime of $\Sigma^{+}$ is long enough so that the 
decay can happen inside the detector. Unlike fermion singlet, $\Sigma$ has gauge interaction and will give 
rich LHC phenomenology. The higgs($\eta$) contribution can be realized if $M_{\eta} \leq M_{\Sigma}$. For 
limiting case, where we have $M_{\eta} \ll M_{\Sigma}$, the value of decay width of the $\Sigma^{0}_{1}$ into 
light doublet$\ell_{L}$ and $\eta$ is $\Gamma = \frac{h^{2}M_{\Sigma}}{8 \pi}$.

Decays of fermion triplet just as those of right handed neutrinos, violate Lepton number. In a machine 
such as LHC, one would typically produce a pair $\Sigma^{+} \Sigma^{0}$ or $\Sigma^{-} \Sigma^{0}$ whose 
decay then allow for interesting $\Delta L=2$ signatures of same sign dileptons and 4 jets.
\begin{figure}[h]
 \centering
 \includegraphics[width=14cm, height=4cm]{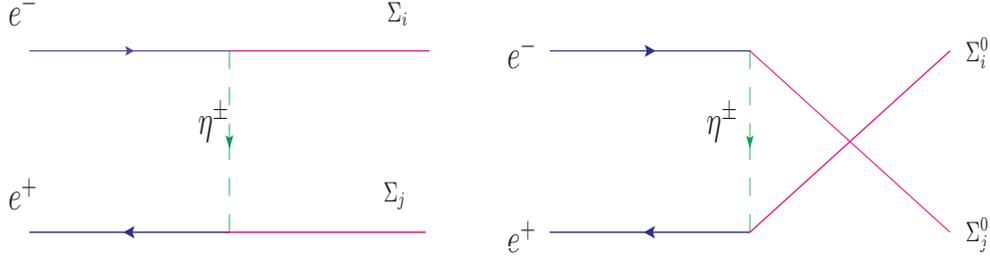}
\caption{Feynman diagrams for annihilation of $\Sigma$ via $\eta$}
\end{figure}

The process like $e^{+}_{}e^{-}_{} \rightarrow \gamma,~ Z,~ \eta \rightarrow \Sigma_{i}^{} \Sigma_{j}^{})$ 
is very important for phenomenological study. If one wants test the production cross-section of fermion 
triplet, then one should know the quantum number of the particle correctly and predict the gauge interaction 
of the particle(fermion triplet) very accurately. The cross-section for this process for TeV scale $\Sigma$ field 
 can be of few $fb$ \cite{strumia2008}.After production of $\Sigma^{0,\pm}$(say), it decays into standard model particles producing 
some jets.
\begin{equation}
\sigma(e^{-} e^{+}\rightarrow \Sigma_{i} \Sigma_{i})= \frac{1}{32 \pi} h^{4}_{ei}(A_{ii}+B_{ii})
\end{equation}
\begin{equation}
\sigma(e^{-} e^{+}\rightarrow \Sigma_{i} \Sigma_{j})= \frac{1}{32 \pi} h^{2}_{ei} h^{2}_{ej} A_{ij}
\end{equation}
where
\begin{eqnarray}
A_{ij}&=&\frac{1}{s(s-4 m^{2}_{e})} \bigg[(2 M^{2}_{\eta^{\pm}}-2 m^{2}_{e}-M^{2}_{\Sigma_{i}}-
M^{2}_{\Sigma_{j}})\times \text{ln}\bigg[\frac{x0-M^{2}_{\eta^{\pm}}}{x1-M^{2}_{\eta^{\pm}}}\bigg]+
\nonumber \\
& &\bigg[\frac{(M^{2}_{\eta^{\pm}}- m^{2}_{e}-M^{2}_{\Sigma_{i}})(M^{2}_{\eta^{\pm}}- m^{2}_{e}-
M^{2}_{\Sigma_{j}})}{(x0-M^{2}_{\eta^{\pm}})(x1-M^{2}_{\eta^{\pm}})} +1\bigg](x0-x1) \bigg]
\end{eqnarray}
and
$$B_{ij}=\frac{M^{2}_{\Sigma_{i}}(s-2 m^{2}_{e})}{s(s-4 m^{2}_{e})
(s-2 m^{2}_{e}-2 M^{2}_{\Sigma_{j}}+2 M^{2}_{\eta^{\pm}})}$$
The value of $x0$ and $x1$ is given by
$$x0=-\frac{1}{4}\left[(s-4 m^{2}_{e})^{\frac{1}{2}}-\frac{s-(M_{\Sigma_{i}}-
M_{\Sigma_{j}})^{2}}{s} \right]^{\frac{1}{2}}$$
$$x1=-\frac{1}{4}\left[(s-4 m^{2}_{e})^{\frac{1}{2}}+\frac{s-(M_{\Sigma_{i}}-
M_{\Sigma_{j}})^{2}}{s} \right]^{\frac{1}{2}}$$
\begin{figure}[ht]
 \begin{center}
 \includegraphics[bb=0 0 240 155]{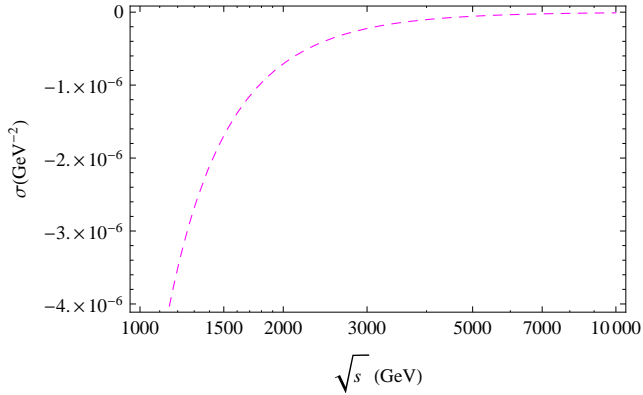}
\end{center}
\caption{Annihilation cross-section for $\Sigma$ with exchange of $\eta^{\pm}$}
\end{figure}
%
\section{Conclusion}
\label{sec:conclusion}
Fermion triplet $(\Sigma^{+},~ \Sigma^{0},~\Sigma^{-})_{i}$(i=1,2,3) gives smallness of neutrino masses 
radiatively and hence many new and interesting possibilities of physics beyond the standard 
model exists. The decay of $\Sigma^{0}_{1}$ into light lepton doublet and an extra Higgs can 
be a source of the matter-antimatter asymmetry of the Univesre. As the scale of particles 
involved in this leptogenesis scenario is of order TeV, then leptogenesis through seesaw can 
be testable at upcoming accelerator(LHC). Phenomenological significance of $\Sigma$ at this 
TeV scale can be verified at LHC. In an accelerator like the LHC, pair production of fermion triplets like 
$pp\rightarrow \Sigma^{+}\Sigma^{0},~\Sigma^{-}\Sigma^{0}$ is possible. Their decay then allow for interesting $\Delta L=2$ signatures. The main point 
is that the triplet fermion is found to be light unlike in the case of right handed fermion 
singlet. After production of a fermion triplet neutrino, its decay into a SM light lepton doublet 
and an extra scalar doublet, i,e, $\Sigma^{0}_{i} \rightarrow \nu_{i} \eta$ which can be tested at upcoming 
accelerator.
\section{Acknowledgments}
I would like to thank Utpal Sarkar, Raghavan Rangarajan  for their valuable comments and for 
reading the manuscript and also like to thank Srubabati Goswami 
and Namit Mahajan for useful discussion. 



\begin{thebibliography}{34}
%
\bibitem{minkowski1977}
P. Minkowski, Phys. Lett. \textbf{67B}, 421 (1977).

\bibitem{yana1979} 
T. Yanagida, in {\it Proc. of the Workshop on Unified Theory and the
Baryon Number of the Universe}, ed. O. Sawada and A. Sugamoto (KEK,
Tsukuba, 1979), p. 95.

\bibitem{grs1979}
M. Gell-Mann, P. Ramond, and R. Slansky, in {\it Supergravity}, ed.
F. van Nieuwenhuizen and D. Freedman (North Holland, Amsterdam,
1979), p. 315.

\bibitem{glashow1980}
S.L. Glashow, in {\it Quarks and Leptons}, ed. M. L$\rm\acute{e}$vy
{\it et al.} (Plenum, New York, 1980), p. 707.

\bibitem{ms1980}
R.N. Mohapatra and G. Senjanovi$\rm\acute{c}$, Phys. Rev. Lett.
\textbf{44}, 912 (1980).

\bibitem{sv1980}
J. Schechter and J.W.F. Valle, Phys. Rev. D \textbf{22}, 2227
(1980).
\bibitem{grimus} W. Grimus, L. Lavoura and B. Radovcic; Phys.Lett.B674:117-121,2009.
\bibitem{lidner} Manfred Lindner and Werner Rodejohann; JHEP 0705:089,2007.
\bibitem{joshi} Robert Foot, H. Lew, X.G. He, Girish C. Joshi; Z.Phys.C44:441,1989.
\bibitem{barr11} S. M. barr; Phys.Rev.Lett.92:101601,(2004). 
\bibitem{barr1}S.M. Barr and Ilja Dorsner; Phys.Lett.B632:527-531,2006. 
\bibitem{kt1990}
E.W. Kolb and M.S. Turner, \textit{The Early Universe},
Addison-Wesley, Reading, MA, 1990.

\bibitem{hambye2004}Thomas Hambye, Yin Lin, Alessio Notari, Michele Papucci, Alessandro Strumia; 
Phys. Lett. B 694(2004) 161-191.
\bibitem{hambye2006} Thomas Hambye, Martti Raidal and Alessandro Strumia; 
Phys. Lett. B. 632 (2006) 667-674.

\bibitem{strumia2008} Roberto Franceschini, Thomas Hambye and Alessandro Strumia; 
Phys. Rev D. 78(2008) 033002.
\bibitem{gu2008}
Pei-Hong Gu and Utpal Sarkar; [arXiv:hep-ph/0811.0956].

\bibitem{wmap2003} 
C.~L.~Bennett {\it et al.},
Astrophys.\ J.\ Suppl.\  {\bf 148} (2003) 1
[arXiv:astro-ph/0302207].

\bibitem{fy1986}
M. Fukugita and T. Yanagida, Phys. Lett. B {\bf 174}, 45 (1986).

\bibitem{ry1986}
P. Langacker, R.D. Peccei, and T. Yanagida, Mod. Phys. Lett. A
\textbf{1}, 541 (1986);

\bibitem{luty1992}
M.A. Luty, Phys. Rev. D \textbf{45}, 455 (1992);

\bibitem{mz1992}
R.N. Mohapatra and X. Zhang, Phys. Rev. D \textbf{46}, 5331 (1992).

\bibitem{fps1995}
M. Flanz, E.A. Paschos, and U. Sarkar, Phys. Lett. B \textbf{345},
248 (1995).

\bibitem{fpsw1996}
M. Flanz, E.A. Paschos, U. Sarkar, and J. Weiss, Phys. Lett. B
\textbf{389}, 693 (1996).


\bibitem{ms1999}
E. Ma and U. Sarkar, Phys. Rev. Lett. \textbf{80}, 5716 (1998).

\bibitem{ma2006}
E. Ma, Phys. Rev. D 73, 077301(2006).

\bibitem{f2008}W. Fischler and R. Flauger, JHEP 09, 020 (2008).

\bibitem{k1985} V. A. Kuzmin, V. A. Rubakov, and M. E. Shaposhnikov, Phys. Lett. B 155, 36 (1985)
\bibitem{m2009}
Ernest Ma: Mod. Phys. Lett. A\,24, 2161(2009) 2491-2495:
arXiv:0908.1770 [hep-ph].
  


\bibitem{mm2009}
E. Ma, Phys. Rev. D 80, 013013 (2009).

\bibitem{ma1998}
E. Ma and U. Sarkar, Phys. Rev. Letts. 80, 5716 (1998).

\bibitem{flantz} M. Flanz, E.A. Paschos, Phys. Rev. D 58 (1998) 11309, Arxiv:hep-ph/9805427.


\bibitem{mm2008}
E. Ma, Phys. Rev. D 78, 017701 (2008).

\bibitem{m2005}
E. Ma, Phys. Lett. B 625, 76 (2005).

\bibitem{mmm2009}
E. Ma and D. Suematsu, Mod. Phys. Lett. A 24, 583 (2009).

\bibitem{di2002}
S. Davidson and A. Ibarra, Phys. Lett. B \textbf{535}, 25 (2002).

\bibitem{bdp2003}
W. Buchm$\rm\ddot{u}$ller, P. Di Bari, and M. Pl$\rm\ddot{u}$macher,
Nucl. Phys. B \textbf{665}, 445 (2003).

\bibitem{sv2006}
A. Strumia and F. Vissani, hep-ph/0606054; and references therein.
\bibitem{reslepto} M. Flanz, E. A. Paschos, U. Sarkar and J. Weiss, Phys.
Lett. B 389, 693 (1996); A. Pilaftsis, Phys. Rev. D 56,
5431 (1997); A. Pilaftsis and T. E. J. Underwood, Nucl.
Phys. B 692, 303 (2004).
\bibitem{hambye}
R. Franceschini, T. Hambye and A. Strumia, Phys. Rev. D 78, 033002(2008).

%
\bibitem{sty} D. Suematsu, T. Toma and Y. Yoshida, Phys. Rev. D79, 093004 (2009).
\bibitem{raidal} E. Ma and M. Raidal, Phys. Rev. Lett. 87, 011802 (2001).
\bibitem{tex} Daijiro Suematsu, Takashi Toma and Tetsuro Yoshida; arXiv:1002.3225 [hep-ph]
\end{thebibliography}
\end{document}